\RequirePackage{amsmath}
\documentclass[runningheads]{llncs}

\newif\iftacas

\tacasfalse

\usepackage[T1]{fontenc}
\usepackage{graphicx}

\PassOptionsToPackage{hyphens}{url}
\usepackage{hyperref}
\hypersetup{
  colorlinks = true,
  citecolor = blue,
  linkcolor = blue,
  urlcolor = blue
}

\usepackage{xcolor}
\usepackage{enumerate}
\usepackage{mathtools}
\usepackage{amsfonts}
\usepackage{amssymb}
\usepackage{nicefrac}
\usepackage{booktabs}
\usepackage{csquotes}
\usepackage{tikz}
\usetikzlibrary{calc}
\usepackage{cite}
\usepackage{pgfplots}
\usepackage{wrapfig}
\pgfplotsset{compat=1.18}

\usepackage{orcidlink}
\renewcommand{\orcidID}[1]{\orcidlink{#1}} 

\usepackage{todonotes}

\definecolor{cola}{rgb}{0.878431, 0.235294, 0.192157}
\definecolor{colb}{rgb}{0.552941, 0.72549, 0.792157}
\definecolor{colc}{rgb}{0.964706, 0.745098, 0}
\definecolor{cold}{rgb}{0.917647, 0.462745, 0}
\definecolor{cole}{rgb}{0.54902, 0.509804, 0.47451}
\definecolor{colf}{rgb}{0.643137, 0.858824, 0.909804}
\definecolor{colg}{rgb}{0.141176, 0.313725, 0.603922}
\definecolor{colh}{rgb}{0.560784, 0.6, 0.243137}

\newcommand{\ComplexityFont}[1]{{\ensuremath{\mathsf{#1}}}}

\newcommand{\NP}{\ComplexityFont{NP}}
\newcommand{\PSPACE}{\ComplexityFont{PSPACE}}
\newcommand{\ETR}{\ComplexityFont{ETR}}

\DeclareMathOperator{\sgn}{sgn}

\newcommand{\step}[2]{\langle{#1},#2\rangle}

\usepackage{rotating} 

\makeatletter

\newcommand*\MY@leftharpoonupfill@{%
    \arrowfill@\leftharpoonup\relbar\relbar
}
\newcommand*\MY@rightharpoonupfill@{%
    \arrowfill@\relbar\relbar\rightharpoonup
}

\newcommand*\overleftharpoon{%
    \mathpalette{\overarrow@\MY@leftharpoonupfill@}%
}
\newcommand*\overrightharpoon{%
    \mathpalette{\overarrow@\MY@rightharpoonupfill@}%
}

\makeatother
\newcommand\harp[1]{\mathstrut\mkern2.5mu#1\mkern-11mu\raise0.6ex%
  \hbox{$\scriptscriptstyle\rightharpoonup$}}

\usepackage{color}

\definecolor{structure}{RGB}{196, 30, 58}

\newcommand{\vertex}[3]{
  \draw[fill=blue] (#1, #2) circle (#3 pt);
}

\newcommand\Overline[2][1pt]{%
    \begin{tikzpicture}[baseline=(a.base)]
      \node[inner xsep=0pt,inner ysep=1.5pt] (a) {$#2$};
      \draw[line width= #1] (a.north west) -- (a.north east);
    \end{tikzpicture}
    }

\newcommand{\aline}[2]{$\Overline{#1#2}$}

\newcommand{\vertexc}[4]{
  \draw[fill=#4] (#1, #2) circle (#3 pt);
}

\newcommand{\orientp}[3]{{\mathsf o}_{#1,#2,#3}}
\newcommand{\orientn}[3]{\overline {{\mathsf o}_{#1,#2,#3}}}

\newcommand{\holep}[3]{{\mathsf h}_{#1,#2,#3}}
\newcommand{\holen}[3]{\overline {{\mathsf h}_{#1,#2,#3}}}

\newcommand{\insidep}[4]{{\mathsf c}_{#1;#2,#3,#4}}
\newcommand{\insiden}[4]{\overline {{\mathsf c}_{#1;#2,#3,#4}}}

\newcommand{\xfour}{{\mathsf u}^4}
\newcommand{\xfive}{{\mathsf u}^5}

\newcommand{\yfour}{{\mathsf v}^4}

\begin{document}

\iftrue
\title{Happy Ending: An Empty Hexagon in~Every~Set~of~30~Points}
\author{Marijn J. H. Heule\inst{1,}\inst{2}\orcidID{0000-0002-5587-8801} \and
Manfred Scheucher\inst{3}\orcidID{0000-0002-1657-9796}}
\authorrunning{M.J.H.~Heule and M.~Scheucher}
\institute{Carnegie Mellon University, Pittsburgh, USA\\
\email{marijn@cmu.edu}\\
\and
Amazon Scholar\\
\and
Institute of Mathematics, Technische Universität Berlin, Germany\\
\email{scheucher@math.tu-berlin.de}}

\else
\title{Happy Ending: An Empty Hexagon
in~Every~Set~of~30~Points}
\fi

\maketitle              
%


\begin{abstract}
Satisfiability solving has been used to tackle a range of long-standing open math problems in recent years. We add another success by solving a geometry problem that originated a century ago. In the 1930s, Esther Klein's exploration of unavoidable shapes in planar point sets in general position showed that every set of five points includes four points in convex position. For a long time, it was open if an empty hexagon, i.e., six points in convex position without a point inside, can be avoided. In 2006, Gerken and Nicolás independently proved that the answer is no. We establish the exact bound: Every 30-point set in the plane in general position contains an empty hexagon. Our key contributions include an effective, compact encoding and a search-space partitioning strategy enabling linear-time speedups even when using thousands of cores.

\end{abstract}

\section{Introduction}
\label{sec:introduction}

In 1932,
Esther Klein showed that every set of five points in the plane \emph{in general position} (i.e., no three points on a common line) 
has a subset of four points in convex position.
Shortly after, Erd\H{o}s and Szekeres \cite{ErdosSzekeres1935} generalized this result by showing that, for every integer $k$, there exists a smallest integer $g(k)$ such that every set of $g(k)$ points in the plane {in general position} 
contains a \emph{$k$-gon} (i.e., a subset of $k$ points
that form the vertices of a convex polygon).
As the research led to the marriage of Szekeres and  Klein,
Erd\H{o}s named it the \emph{happy ending problem}.
Erd\H{o}s and Szekeres constructed witnesses of $g(k) > 2^{k-2}$ \cite{ErdosSzekeres1960},
which they conjectured to be maximal.
The best upper bound is $g(k) \le 2^{k+o(k)}$ \cite{Suk2017,HolmsenMPT2020}.

Determining the value $g(5)=9$ requires a more involved case distinction compared to $g(4)=5$\cite{KalbfleischKalbfleischStanton1970}.
It took until 2006 to determine that  $g(6)=17$ via an exhaustive computer search by Szekeres and Peters \cite{SzekeresPeters2006} using
 1500 CPU hours.
Marić \cite{Maric2019} and
Scheucher \cite{Scheucher2020} 
independently verified $g(6) = 17$ using satisfiability (SAT) solving in a few CPU hours.
This was later reduced to 10 CPU minutes~\cite{Scheucher2023}.
The approach presented in this paper computes it in 8.53 CPU seconds,
showing the effectiveness of SAT compared to the original method.

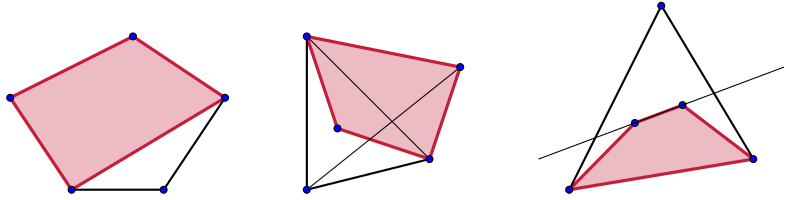
\begin{figure}[t]
    \centering
    \newcommand{\scale}{1.7}
    \newcommand{\x}{0.3}

\hbox{}
\hfill
\begin{tikzpicture}[scale=\scale]

\tikzset{vertex/.style={draw, circle, fill=blue, scale=\x}}

\filldraw[draw=structure, very thick, fill=structure!30!white]
(2.68,3.04)
 -- (1.96,3.52)
 -- (1,3.04)
 -- (1.48,2.32)
 -- cycle;     

\node[vertex] (a) at (2.2,2.32) {};
\node[vertex] (b) at (2.68,3.04) {};
\node[vertex] (c) at (1.96,3.52) {};
\node[vertex] (d) at (1,3.04) {};
\node[vertex] (e) at (1.48,2.32) {};

\draw[thick] (e) -- (a) -- (b);

\end{tikzpicture}
\hfill
\begin{tikzpicture}[scale=\scale]

\tikzset{vertex/.style={draw, circle, fill=blue, scale=\x}}

\filldraw[draw=structure, very thick, fill=structure!30!white]
 (4.04,2.8)
 -- (3.8,3.52)
 -- (5,3.28)
 -- (4.76,2.56)
 -- cycle;     

\node[vertex] (a) at (4.04,2.8) {};
\node[vertex] (b) at (3.8,3.52) {};
\node[vertex] (c) at (5,3.28) {};
\node[vertex] (d) at (4.76,2.56) {};
\node[vertex] (e) at (3.8,2.32) {};

\draw (e) -- (c) (d) -- (b);
\draw[thick] (b) -- (e) -- (d);

\end{tikzpicture}
\hfill
\begin{tikzpicture}[scale=\scale]

\tikzset{vertex/.style={draw, circle, fill=blue, scale=\x}}

\filldraw[draw=structure, very thick, fill=structure!30!white]
(6.84656,2.98246)
 -- (6.4739,2.842713)
 -- (5.96,2.32)
 -- (7.4,2.56)
 -- cycle;     

\node[vertex] (a) at (6.4739,2.842713) {};
\node[vertex] (b) at (6.84656,2.98246) {};
\node[vertex] (c) at (5.96,2.32) {};
\node[vertex] (d) at (6.68,3.76) {};
\node[vertex] (e) at (7.4,2.56) {};

\draw[thick] (c) -- (d) -- (e);

\draw[shift={(5.72, 2.56)}, scale=1.5]
(0, 0)
 -- (1.28, 0.48);

\end{tikzpicture}
\hfill
\hbox{}
    \caption{An illustration for the proof of $h(4)= 5$: The three possibilities of how five points can be placed. Each possibility implies a $4$-hole.}
    \label{fig:h4is5}
\end{figure}

Erd\H{o}s also asked 
whether every sufficiently large point set contains a \emph{$k$-hole}:
a $k$-gon without a point inside. 
We denote by $h(k)$ the smallest integer---if it exists---such
that every set of $h(k)$ points in general position in the plane contains a $k$-hole.
Both $h(3)=3$ and $h(4)=5$ are easy to compute (see Fig.~\ref{fig:h4is5} for an illustration) 
and coincide with the original setting.
Yet the answer can differ a lot,
as Horton~\cite{Horton1983} constructed arbitrarily large point sets
without 7-holes.

While Harborth \cite{Harborth1978} 
showed in 1978 that $h(5)=10$,
the existence of $6$-holes remained open until the late 2000s,
when Gerken \cite{Gerken2008}\footnote{Gerken's groundbreaking work was awarded the Richard-Rado prize by the German Mathematical Society in 2008.} and Nicol\'as \cite{Nicolas2007} independently proved that $h(6)$ is finite.
Gerken proved that every $9$-gon yields a $6$-hole, thereby showing that $h(6)\le g(9) \le 1717$~\cite{TothValtr2004}.
The best-known lower bound $h(6) \ge 30$ is witnessed by a set of 29 points without $6$-holes 
which was found by Overmars \cite{Overmars2002} 
using a local search
\iftacas
approach. 
\else
approach, see Figure~\ref{fig:overmars}. 
\fi

We close the gap between the upper and lower bound and ultimately answer Erd\H{o}s' question by proving that every set of 30 points yields a 6-hole.
\begin{theorem}
\label{thm:h6=30}
$h(6) = 30$. 
\end{theorem}
Our result is actually stronger and shows that the bounds for $6$-holes in point sets coincide with the bounds for $6$-holes in 
\emph{counterclockwise systems}~\cite{Knuth1992}. 
This represents another success of solving long-standing open problems in mathematics using SAT, similar to results on Schur Number Five~\cite{Schur} and Keller's Conjecture~\cite{Keller}. 

We also investigate 
the combination of $6$-holes and $7$-gons and 
show 
\begin{theorem}
\label{thm:h6g7=24}
Every set of 24 points in the plane in general position contains a $6$-hole or a $7$-gon.
\end{theorem}

\paragraph{}
We achieve these results through the following contributions:
\begin{itemize}
\item We develop a compact and effective SAT encoding for $k$-gon and $k$-hole problems that uses $O(n^4)$ clauses, while existing 
encodings use $O(n^k)$ clauses.
\item We construct a partitioning of $k$-gon and $k$-hole problems that allows us to solve them with linear-time speedups even when using thousands of cores.
\item We present a novel method of validating SAT-solving results that checks the proof while solving the problem using substantially less overhead.  
\item We verify most of the presented results using clausal proof checking.
\end{itemize}

\section{Preliminaries}
\label{sec:preliminaries}

\paragraph*{The SAT problem.}
The satisfiability problem
(SAT) asks whether a Boolean formula can be satisfied by some assignment of truth values to its variables. 
The Handbook of Satisfiability~\cite{HandbookSatisfiablity2009} provides an overview.
%
We consider
formulas in \emph{conjunctive normal form} (CNF), which is the default input of SAT solvers. 
As such, a formula $\Gamma$ is a conjunction (logical ``AND'') of {\em clauses}. A clause is a disjunction (logical ``OR'') of literals, where a literal is a Boolean variable or its negation. 
We sometimes write (sets of) clauses using other logical connectives. 

If a formula $\Gamma$ is found to be satisfiable, modern SAT solvers commonly output a truth assignment of the variables.
Additionally, if a formula turns out to be unsatisfiable, sequential SAT solvers produce an independently-checkable proof that there exists no assignment that satisfies the formula.

\paragraph*{Verification.}
The most commonly-used proofs for SAT problems are expressed in the DRAT clausal proof system~\cite{Heule2016}.
A DRAT proof of unsatisfiability is a list of clause addition and clause deletion steps.
Formally, a clausal proof is a list of pairs $\step{s_1}{C_1},\dots,\step{s_m}{C_m}$, where for each $i\in \{ 1,\dots,m\}$, $s_i \in \{\mathsf{a}, \mathsf{d}\}$ and $C_i$ is a clause.
If $s_i = \mathsf{a}$, the pair is called an \emph{addition}, and if $s_i = \mathsf{d}$, it is called a \emph{deletion}.
For a given input formula $\Gamma_0$, a clausal proof gives rise to a set of \emph{accumulated formulas} $\Gamma_i$ ($i \in \{1,\dots,m\}$) as follows:
\begin{align*}
\Gamma_i = 
\begin{cases}
\Gamma_{i-1} \cup \{C_i\}	& \text{if $\mathsf{s}_i = \mathsf{a}$}\\
\Gamma_{i-1} \setminus \{C_i\}	& \text{if $\mathsf{s}_i = \mathsf{d}$}\\
\end{cases}
\end{align*}

Each clause addition must preserve satisfiability, which is usually guaranteed by requiring the added clauses to fulfill some efficiently decidable syntactic criterion.
Deletions help to speed up proof checking by keeping the accumulated formula small.
A valid proof of unsatisfiability must add the empty clause.

\paragraph*{Cube And Conquer.}
The cube-and-conquer approach~\cite{HKWB2012_cubes} aims to \emph{split} a SAT instance  $\Gamma$ into multiple instances $\Gamma_1, \ldots, \Gamma_m$ in such a way that $\Gamma$ 
is satisfiable if and only if at least one of the instances $\Gamma_i$ is satisfiable, thus allowing work on the
different instances $\Gamma_i$  in parallel.
A {\em cube} is a conjunction of literals. 
Let $\psi = \left(c_1\lor \cdots \lor c_m\right)$ be a disjunction of cubes. When $\psi$ is a tautology, we have
\[
\Gamma \iff \Gamma \land \psi \iff \bigvee_{i=1}^m (\Gamma \land c_i) \iff \bigvee_{i=1}^m \Gamma_i ,
\]
where the different $\Gamma_i \coloneqq (\Gamma \land c_i)$ are the instances resulting from the split.

Intuitively, each cube $c_i$ represents a \emph{case}, i.e., an assumption about a satisfying assignment to $\Gamma$, and soundness comes from $\psi$ being a tautology, which means that the split into cases is exhaustive. If the split is well designed, then each $\Gamma_i$ is a particular case that is substantially easier to solve than $\Gamma$, and thus solving them all in parallel can give significant speed-ups, especially considering the sequential nature of CDCL at the core of most solvers.

However, the quality of the split ($\psi$) has an enormous impact on the effectiveness of the approach. A key challenge is figuring out a high-quality split.

\section{Trusted Encoding}
\label{sec:encoding}

To obtain an upper-bound result using a SAT-based approach, we need to show that every set of $n$ points contains a $k$-hole. We will do this by constructing a formula based on $n$ points that asks whether a $k$-hole can be avoided. If this formula is unsatisfiable, then we obtain the bound $h(k) \leq n$. Instead of reasoning directly whether an empty $k$-gon can be avoided, we ask whether every $k$ points contain at least one triangle with a point inside. The latter implies the former.

\begin{figure}[t]

\begin{minipage}{.24\textwidth}
\begin{tikzpicture}
\node at (-0.25,-0.1) {$a$};
\node at (1.55,-0.1) {$b$};
\node at (1.8,1.4) {$c$};
\node at (1,0.15) {$i$};
\node[draw, circle, fill=blue, scale=0.3] (a) at (0,0) {};
\node[draw, circle, fill=blue, scale=0.3] (b) at (1.25,0) {};
\node[draw, circle, fill=blue, scale=0.3] (c) at (2,1.25) {};
\node[draw, circle, fill=blue, scale=0.3] (i) at (1,0.4) {};
\draw[thick] (a) -- (b) -- (c) -- (a);
\draw[-latex] (0.7,0) -- (0.7,0.3);
\draw[-latex] (1.2,0.75) -- (1.35,0.5);
\end{tikzpicture}
\end{minipage}
\hfil
\begin{minipage}{.24\textwidth}
\begin{tikzpicture}
\node at (-0.25,0.1) {$a$};
\node at (1.55,0.1) {$b$};
\node at (1.8,-1.4) {$c$};
\node at (1,-0.15) {$i$};
\node[draw, circle, fill=blue, scale=0.3] (a) at (0,0) {};
\node[draw, circle, fill=blue, scale=0.3] (b) at (1.25,0) {};
\node[draw, circle, fill=blue, scale=0.3] (c) at (2,-1.25) {};
\node[draw, circle, fill=blue, scale=0.3] (i) at (1,-0.4) {};
\draw[thick] (a) -- (b) -- (c) -- (a);
\draw[-latex] (0.7,0) -- (0.7,-0.3);
\draw[-latex] (1.2,-0.75) -- (1.35,-0.5);
\end{tikzpicture}
\end{minipage}
\hfil
\begin{minipage}{.24\textwidth}
\begin{tikzpicture}
\node at (-0.25,-0.1) {$a$};
\node at (1.55,-0.1) {$b$};
\node at (1.8,1.4) {$c$};
\node at (1.3,0.6) {$i$};
\node[draw, circle, fill=blue, scale=0.3] (a) at (0,0) {};
\node[draw, circle, fill=blue, scale=0.3] (b) at (1.25,0) {};
\node[draw, circle, fill=blue, scale=0.3] (c) at (2,1.25) {};
\node[draw, circle, fill=blue, scale=0.3] (i) at (1.5,0.8) {};
\draw[thick] (a) -- (b) -- (c) -- (a);
\draw[-latex] (0.8,0.5) -- (0.95,0.25);
\draw[-latex] (1.40,0.25) -- (1.15,0.4);
\end{tikzpicture}
\end{minipage}
\hfil
\begin{minipage}{.24\textwidth}
\begin{tikzpicture}
\node at (-0.25,0.1) {$a$};
\node at (1.55,0.1) {$b$};
\node at (1.8,-1.4) {$c$};
\node at (1.3,-0.6) {$i$};
\node[draw, circle, fill=blue, scale=0.3] (a) at (0,-0) {};
\node[draw, circle, fill=blue, scale=0.3] (b) at (1.25,-0) {};
\node[draw, circle, fill=blue, scale=0.3] (c) at (2,-1.25) {};
\node[draw, circle, fill=blue, scale=0.3] (i) at (1.5,-0.8) {};
\draw[thick] (a) -- (b) -- (c) -- (a);
\draw[-latex] (0.8,-0.5) -- (0.95,-0.25);
\draw[-latex] (1.40,-0.25) -- (1.15,-0.4);
\end{tikzpicture}
\end{minipage}

\caption{The four ways a point $p_i$ can be inside triangle $\{p_a,p_b,p_c\}$ based on whether
$i < b$ (left two images) and whether $p_c$ is above the line $p_ap_b$ (first and third image).
}
\label{fig:inside}
\end{figure}
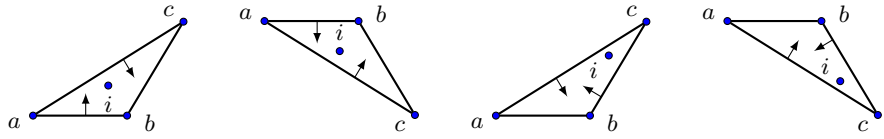

We only need to know for each triple of points whether it is empty. Throughout the paper, we assume that points are sorted with strictly increasing $x$-coordinates. This gives us only four options for a point $p_i$ to be inside the triangle formed by points $p_a$, $p_b$, $p_c$, see Fig.~\ref{fig:inside}.
For example, the left image shows that $p_i$ is inside if $a < i < b$, $p_c$ and $p_i$ are above the line \aline{p_a}{p_b}, and $p_i$ is below the line \aline{p_a}{p_c}.
So we need some machinery to express that points are above or below certain lines. That is what the encoding will provide. 
For readability, we sometimes identify points by their indices, that is, we refer to $p_a$ by its index~$a$.

We first present what we call the \emph{trusted encoding} to determine whether a $6$-hole can be avoided. 
The encoding needs to be trusted in the sense that we do not provide a mechanically verified proof of its correctness.
Building upon existing work~\cite{Scheucher2020}, our primary focus is on $6$-holes, which constitute our main result.
The encoding of $6$-gons and $7$-gons is similar and more simple.
During an initial study, the estimated runtime for showing $h(6) \leq 30$ using this encoding and off-the-shelf partitioning was roughly 1000 CPU years. 
The optimizations in Sections~\ref{sec:optimization} and~\ref{sec:partitioning} reduce the computational costs to about 2 CPU years. 

\subsection{Orientation Variables}
\label{sec:axiom}

\begingroup
\setlength{\intextsep}{3pt}%
\setlength{\columnsep}{10pt}%

\begin{wrapfigure}{r}{3.3cm}
    \centering
    \vspace{-10pt}
    \begin{tikzpicture}
    \newcommand{\x}{1.5}
    
    \draw[-,thick] (-0.25,0.05) -- (2.75,-0.55); 
    
    \vertex{0}{0}{\x}
    \vertex{1}{-0.2}{\x}
    
    \node at (-0.1,-0.3) {$a$};
    \node at (0.9,-0.5) {$b$};
    
    \vertexc{2}{0.4}{\x}{green!50!black};
    \vertexc{2.5}{-1.5}{\x}{structure};

    \node at (1.8,0.6) {$c$};
    \node at (2.2,-1.6) {$d$};
    
    \draw[-latex,thick,color=green!50!black] (0.2,0.3) to [out=345,in=250] (2.5,0.4);
    \draw[-latex,thick,color=structure] (0.1,-0.8) to [out=0,in=-250] (2.75,-1.6);

    \node at (1.3,0.3) {\textcolor{green!50!black}{$\boldsymbol{+}$}};
    \node at (1.6,-1.1) {\textcolor{structure}{\bf $\boldsymbol{-}$}};
    
    \end{tikzpicture}
        \vspace{0pt}

    \caption{An illustration of triple orientations.}
    \label{fig:threepoints}
\end{wrapfigure}
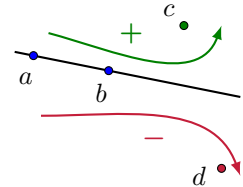

We formulate the problem in such a way that all reasoning is based solely on the relative positions of points.
Thus, we do not encode coordinates but only orientations of point triples.
For a point set $S=\{p_1,\ldots,p_n\}$ with $p_i = (x_i,y_i)$, 
the triple $(p_a,p_b,p_c)$ with $a<b<c$
is \emph{positively oriented} (resp.\ \emph{negatively oriented})
if $p_c$ lies above (resp.\ below) 
the line \aline{p_a}{p_b} through 
$p_a$ and~$p_b$. 
The notion of positive orientation corresponds to Knuth's \emph{counterclockwise relation}~\cite{Knuth1992}. 
Fig.~\ref{fig:threepoints} illustrates a po\-si\-tive\-ly-o\-ri\-en\-ted triple $(p_a,p_b,p_c)$ and a ne\-ga\-tive\-ly-o\-ri\-en\-ted triple $(p_a,p_b,p_d)$.

To search for point sets without $k$-gons and $k$-holes, we introduce a Boolean {\em orientation variable}
$\orientp{a}{b}{c}$ for each triple $(p_a,p_b,p_c)$ with $a<b<c$. 
Intuitively, $\orientp{a}{b}{c}$ is supposed to be true if the triple is positively oriented. 
Since we assume general position, no three points lie on a common line, 
so $\orientp{a}{b}{c}$ being false means that the triple is negatively oriented.

\subsection{Containment Variables, $3$-Hole Variables, and Constraints}

Using orientation variables, we can now express what it means for a triangle to be empty.
We define \emph{containment variables} $\insidep{i}{a}{b}{c}$ to encode whether point~$p_i$ lies inside the triangle spanned by $\{p_a,p_b,p_c\}$.
Since the points have increasing $x$-coordinates, containment is only possible if $a<i<c$.
We use two kinds of definitions, depending on whether $i$ is smaller or larger than~$b$ (see Fig.~\ref{fig:inside}). 
The first definition is for the case $a < i < b$. Note that if $\orientp{a}{b}{c}$ is true, we only need to know whether
$i$ is above the line \aline{p_a}{p_b} and below the line \aline{p_a}{p_c}. Earlier work~\cite{Scheucher2020} used an extended definition that included the redundant variable $\orientp{i}{b}{c}$. Avoiding this variable makes the definition more compact (six instead of eight clauses) and the resulting formula is easier to solve.
\begin{equation}
\insidep{i}{a}{b}{c} \leftrightarrow \Big(\big(\orientp{a}{b}{c} \rightarrow (\orientn{a}{i}{b} \land \orientp{a}{i}{c})\big) \land \big(\orientn{a}{b}{c} \rightarrow (\orientp{a}{i}{b} \land \orientn{a}{i}{c})\big)\Big)
\label{eq:inside1}
\end{equation}
The second definition is for $b < i < c$, which avoids using the variable $\orientp{a}{b}{i}$:
\begin{equation}
\insidep{i}{a}{b}{c} \leftrightarrow \Big(\big(\orientp{a}{b}{c} \rightarrow (\orientp{a}{i}{c} \land \orientn{b}{i}{c})\big) \land \big(\orientn{a}{b}{c} \rightarrow (\orientn{a}{i}{c} \land \orientp{b}{i}{c})\big)\Big)
\label{eq:inside2}
\end{equation}
Each definition translates into six clauses (without using Tseitin variables). 

Additionally, we introduce definitions $\holep{a}{b}{c}$ of \emph{$3$-hole variables} that express whether the triangle spanned by $\{p_a,p_b,p_c\}$ is a $3$-hole. 
The triangle $\{p_a,p_b,p_c\}$ forms a $3$-hole if and only if no point $p_i$ lies in its interior.
A point $p_i$ can only be an inner point if it lies in the vertical strip between $p_a$ and~$p_c$ and if it is distinct from~$p_b$.
Since the points are sorted, the index $i$ of an interior point $p_i$ must therefore fulfill $a<i<c$ and $i \neq b$. 
Logically, the definition is as follows:
\begin{equation}
\holep{a}{b}{c} \leftrightarrow \bigwedge_{\substack{a<i<c\\ i \neq b}} \insiden{i}{a}{b}{c}.
\label{eq:hole}
\end{equation}

Finally, we encode the ``forbid $k$-hole'' constraint as follows: For each subset $X \subseteq S$ of size $k$, at least one of the triangles formed by three points in $X$ must not be a $3$-hole. So for $k=6$, each clause consists of $\binom{k}{3}=20$ literals.

\begin{equation}
\label{eq:holes_constraints}
\bigwedge_{\substack{X \subseteq S\\ |X| = k}}
~~\big(~
\bigvee_{\substack{a,b,c \in X\\ a<b<c}} 
\holen{a}{b}{c} ~\big)  
\end{equation}

In Section~\ref{sec:optimization}, we will optimize the encoding. Most optimizations aim to improve the encoding of the constraint (\ref{eq:holes_constraints}).

\subsection{Forbidding Non-Realizable Patterns}

Only a small fraction of all assignments to the $\binom{n}{3}$ orientation variables, $2^{\Theta(n \log n)}$, actually describe point sets~\cite{BjoenerLVWSZ1993}.
However, we can reduce the search space from $2^{\Theta(n^3)}$ to $2^{\Theta(n^2)}$ by forbidding non-realizable patterns~\cite{Knuth1992}. 
Consider four points $p_a,p_b,p_c,p_d$ in a sorted point set with $a<b<c<d$.
The leftmost three points 
determine three lines \aline{p_a}{p_b}, \aline{p_a}{p_c}, \aline{p_b}{p_c}, 
which partition the open half-plane $\{(x,y) \in \mathbb{R}^2 : x>x_c\}$  into four regions (see Fig.~\ref{fig:fourpoints}). After placing $p_a$, $p_b$, $p_c$, observe that all realizable positions of point $p_d$ obey the following implications:
$\orientp{a}{b}{c} \land \orientp{a}{c}{d} \Rightarrow \orientp{a}{b}{d}$ and $\orientp{a}{b}{c} \land \orientp{b}{c}{d} \Rightarrow \orientp{a}{c}{d}$.
Similarly for the negations, $\orientn{a}{b}{c} \land \orientn{a}{c}{d} \Rightarrow \orientn{a}{b}{d}$ and $\orientn{a}{b}{c} \land \orientn{b}{c}{d} \Rightarrow \orientn{a}{c}{d}$. These implications are equivalent to the following clauses (grouping positive and negative):
\begin{eqnarray}
(\orientn{a}{b}{c} \lor \orientn{a}{c}{d} \lor \orientp{a}{b}{d}) 
&\land& (\orientp{a}{b}{c} \lor \orientp{a}{c}{d} \lor \orientn{a}{b}{d}) \label{eq:axiom1}\\
(\orientn{a}{b}{c} \lor \orientn{b}{c}{d} \lor \orientp{a}{c}{d})
&\land& (\orientp{a}{b}{c} \lor \orientp{b}{c}{d} \lor \orientn{a}{c}{d}) \label{eq:axiom2}
\end{eqnarray}

Forbidding these non-realizable assignments was also used for $g(6) \leq 17$~\cite{SzekeresPeters2006}. Some call the restriction {\em signotope axioms}~\cite{FelsnerWeil2001}.
The counterclockwise system axioms~\cite{Knuth1992} achieve the same effect, but require $\Theta(n^5)$ clauses instead of $\Theta(n^4)$.

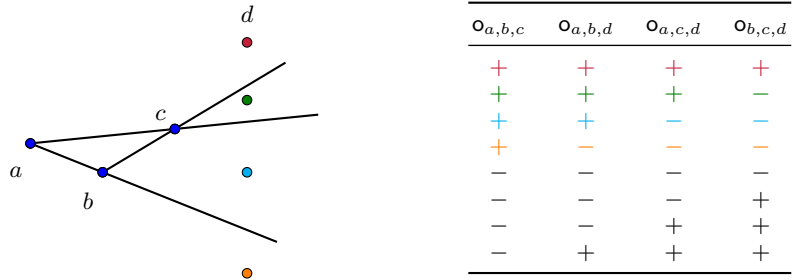
\begin{figure}[t]
    \centering
        \begin{minipage}{.5\textwidth}
\begin{tikzpicture}[scale=1.2]
\newcommand{\x}{1.5}
\vertex{.32}{6.56}{\x};
\vertex{1.12}{6.24}{\x};
\vertex{1.92}{6.72}{\x};
  \draw[shift={(.32, 6.56)}, scale=3.414, thick,-]
    (0, 0)
     -- (.80, -.32);
  \draw[shift={(1.12, 6.24001)}, scale=2.5322, thick, -]
    (0, 0)
     -- (.80, .48);
  \draw[shift={(3.5081, 6.8788)}, scale=1.9926, thick, -]
    (0, 0)
     -- (-1.60, -.16);
\vertexc{2.72}{5.12}{\x}{orange};
\vertexc{2.72}{6.24}{\x}{cyan};
\vertexc{2.72}{7.04}{\x}{green!50!black};
\vertexc{2.72}{7.68}{\x}{structure};
  \node
     at (.16, 6.24) {$a$};
  \node
     at (.96, 5.92) {$b$};
  \node
     at (1.76, 6.88) {$c$};
  \node
     at (2.72, 8) {$d$};

\vertex{.32}{6.56}{\x};
\vertex{1.12}{6.24}{\x};
\vertex{1.92}{6.72}{\x};
\end{tikzpicture}

\end{minipage}
    \begin{minipage}{.4\textwidth}

\begin{tabular}{c@{~~~~}c@{~~~~}c@{~~~~}c}
\toprule
$\orientp{a}{b}{c}$ & $\orientp{a}{b}{d}$ & $\orientp{a}{c}{d}$ & $\orientp{b}{c}{d}$\\
\midrule
\textcolor{structure}{$+$} & \textcolor{structure}{$+$} & \textcolor{structure}{$+$} & \textcolor{structure}{$+$}\\[-1pt]
\textcolor{green!50!black}{$+$}      & \textcolor{green!50!black}{$+$}      & \textcolor{green!50!black}{$+$}      & \textcolor{green!50!black}{$-$}\\[-1pt]
\textcolor{cyan}{$+$}        & \textcolor{cyan}{$+$}       & \textcolor{cyan}{$-$}         & \textcolor{cyan}{$-$}\\[-1pt]
\textcolor{orange}{$+$}    & \textcolor{orange}{$-$}     & \textcolor{orange}{$-$}     & \textcolor{orange}{$-$}\\[-1pt]

$-$ & $-$ & $-$ & $-$\\[-1pt]
$-$ & $-$ & $-$ & $+$\\[-1pt]
$-$ & $-$ & $+$ & $+$\\[-1pt]
$-$ & $+$ & $+$ & $+$\\
\bottomrule
\end{tabular}
\end{minipage}
    \caption{All possibilities to place four points, when points are sorted from left to right.}
    \label{fig:fourpoints}
\end{figure}

\subsection{Initial Symmetry Breaking}
\label{sec:initial_symmetry_breaking}

To further reduce the search space, we ensure that $p_1$ lies on the boundary of the convex hull (i.e., it is an extremal point) 
and that $p_2,\ldots,p_n$ appear around $p_1$ in counterclockwise order, thus providing us the unit clauses  $(\orientp{1}{a}{b})$ for $1<a<b$.
Without loss of generality, we can label points to satisfy the above, because the labeling doesn't affect gons and holes. However, we also want
points to be sorted from left to right. One can satisfy both orderings at the same time using the lemma below.
We attach a proof in 
\iftacas
the extended version~\cite{HeuleScheucher2024extended}.
\else
Appendix~\ref{app:proof_of_lemma}.
\fi

\endgroup

\begin{lemma}[\!\!{\cite[Lemma~1]{Scheucher2020}}]
\label{lemma:increasing_coordinates}
Let $S=\{p_1,\ldots,p_n\}$ 
be a point set in the plane 
in general position such that
$p_1$ is extremal and 
$p_2,\ldots,p_n$ appear (clockwise or counterclockwise) around~$p_1$.
Then there exists a point set 
$\tilde{S}=\{\tilde{p}_1,\ldots,\tilde{p}_n\}$ with the same triple orientations
(in particular, $\tilde{p}_1$ is extremal and 
$\tilde{p}_2,\ldots,\tilde{p}_n$ appear around~$\tilde{p}_1$)
such that the points $\tilde{p}_1,\ldots,\tilde{p}_n$ have increasing $x$-coordinates.
\end{lemma}

\section{Optimizing the Encoding}
\label{sec:optimization}

An ideal SAT encoding has the following three properties: 
\begin{enumerate}[1)]
    \item 
    it is compact to reduce the cost of unit propagation (and cache misses); 
    
    \item 
    it detects conflicts as early as possible (i.e., is domain consistent~\cite{Gent2002ArcCI}); and 

    \item 
    it contains variables that can generalize conflicts effectively. 
\end{enumerate}
\goodbreak

The trusted encoding lacks these properties because it has $O(n^6)$ clauses, cannot quickly detect holes, and has no variables that can generalize conflicts. In this section, we show how to modify the trusted encoding to obtain all three properties. All the modifications are expressible in a proof to ensure correctness.

\subsection{Toward Domain Consistency}
\label{sec:arc}

The effectiveness of an encoding depends on how quickly the solver can 
determine a conflict. Given an assignment, we want to derive as much as
possible via unit propagation. This is known as \emph{domain consistency}~\cite{Gent2002ArcCI}. The trusted encoding does not have this property.
We modify the encoding below to boost propagation. 

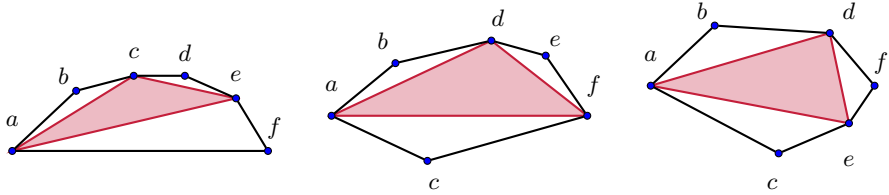
\begin{figure}[tb]

\begin{tikzpicture}[xscale=0.85]
\node at (0,0.4) {$a$};
\node at (0.8,1) {$b$};
\node at (1.9,1.3) {$c$};
\node at (2.7,1.3) {$d$};
\node at (3.5,1) {$e$};
\node at (4.1,0.3) {$f$};

\filldraw[draw=structure, thick, fill=structure!30!white]
(0,0) -- (1.9,1) -- (3.5,0.7) -- cycle;

\node[draw, circle, fill=blue, scale=0.3] (a) at (0,0) {};
\node[draw, circle, fill=blue, scale=0.3] (b) at (1,0.8) {};
\node[draw, circle, fill=blue, scale=0.3] (c) at (1.9,1) {};
\node[draw, circle, fill=blue, scale=0.3] (d) at (2.7,1) {};
\node[draw, circle, fill=blue, scale=0.3] (e) at (3.5,0.7) {};
\node[draw, circle, fill=blue, scale=0.3] (f) at (4,0) {};

\draw[black, thick] (a) -- (b) -- (c) -- (d) -- (e) -- (f) -- (a);
\node at (0,-0.5) {~~};
\end{tikzpicture}
\hfil
\begin{tikzpicture}[xscale=0.85]
\node at (0,0.4) {$a$};
\node at (0.8,1) {$b$};
\node at (1.6,-0.9) {$c$};
\node at (2.6,1.3) {$d$};
\node at (3.5,1) {$e$};
\node at (4.1,0.4) {$f$};

\filldraw[draw=structure, thick, fill=structure!30!white]
(0,0) -- (2.5,1) -- (4,0) -- cycle;

\node[draw,circle, fill=blue, scale=0.3] (a) at (0,0) {};
\node[draw,circle, fill=blue, scale=0.3] (b) at (1,0.7) {};
\node[draw,circle, fill=blue, scale=0.3] (c) at (1.5,-0.6) {};
\node[draw,circle, fill=blue, scale=0.3] (d) at (2.5,1) {};
\node[draw,circle, fill=blue, scale=0.3] (e) at (3.35,0.8) {};
\node[draw,circle, fill=blue, scale=0.3] (f) at (4,0) {};
\draw[black, thick] (a) -- (b) -- (d) -- (e) -- (f) -- (c) -- (a);
\end{tikzpicture}
\hfil
\begin{tikzpicture}[xscale=0.85]
\node at (0,0.4) {$a$};
\node at (0.8,1) {$b$};
\node at (1.9,-1.3) {$c$};
\node at (3.1,1) {$d$};
\node at (3.1,-1) {$e$};
\node at (3.6,0.3) {$f$};

\filldraw[draw=structure, thick, fill=structure!30!white]
(0,0) -- (2.8,0.7) -- (3.1,-0.5) -- cycle;

\node[draw,circle, fill=blue, scale=0.3] (a) at (0,0) {};
\node[draw,circle, fill=blue, scale=0.3] (b) at (1,0.8) {};
\node[draw,circle, fill=blue, scale=0.3] (c) at (2,-0.9) {};
\node[draw,circle, fill=blue, scale=0.3] (d) at (2.8,0.7) {};
\node[draw,circle, fill=blue, scale=0.3] (e) at (3.1,-0.5) {};
\node[draw,circle, fill=blue, scale=0.3] (f) at (3.5,0) {};
\draw[black, thick] (a) -- (b) -- (d) -- (f) -- (e) -- (c) -- (a);
\end{tikzpicture}
    
    \vspace{-10pt}
    \caption{Three types of $6$-gons: left,~all points are on one side of line \aline{a}{\mathit{f}} (2 cases); 
    middle,~three points are on one side and one point is on the other side of line \aline{a}{\mathit{f}} (8~cases); 
    and right,~two points are on either side of line \aline{a}{\mathit{f}} (6 cases). If the marked triangle is empty,
    we can conclude that there exists a $6$-hole.}
    \label{fig:holes}
\end{figure}

We borrow from the method by Szekeres and Peters that a $k$-gon can be detected by looking at assignments to $k-2$ orientation variables~\cite{SzekeresPeters2006}. For example, if $\orientp{a}{b}{c}$, $\orientp{b}{c}{d}$, $\orientp{c}{d}{e}$, and $\orientp{d}{e}{f}$ with 
$a\!<\!b\!<\!c\!<\!d\!<\!e\!<\!f$ are assigned to the same truth value, then this implies that the points form a $6$-gon. 
An illustration of this assignment is shown in Fig.~\ref{fig:holes} (left).
We combine this with our observation below that only a specific triangle has to be empty to infer a $6$-hole somewhere. 

Consider a scenario involving six points, $a$, $b$, $c$, $d$, $e$, and $f$, that are arranged from left to right. In this scenario, the orientation variables $\orientp{a}{b}{c}$, $\orientp{b}{c}{d}$, $\orientp{c}{d}{e}$, and $\orientp{d}{e}{f}$ are all set to false, while the $3$-hole variable $\holep{a}{c}{e}$ is set to true. As mentioned above, this implies that the points form a $6$-gon. Together with $3$-hole variable $\holep{a}{c}{e}$ being set to true, we can deduce the existence of a $6$-hole: The $6$-gon is either a $6$-hole or it contains a $6$-hole. The reasoning will be explained in the next paragraph. Note that in the trusted encoding of this scenario, only one out of the twenty literals in the corresponding `forbid $6$-hole' clause is false. This suggests that the solver is still quite far from detecting a conflict.

A crucial insight underpinning our efficient encoding is the understanding that the truth of the variable $\holep{a}{c}{e}$ alone is sufficient to infer the existence of a $6$-hole. Consider the following rationale: If the triangle $\{a, b, c\}$ contains any points, then there must be at least one point inside the triangle that is closer to the line \aline{a}{c} than point $b$ is. Let's denote the nearest point as $i$. The proximity of $i$ to the line \aline{a}{c} guarantees that the triangle $\{a, i, c\}$ is empty. We can substitute $b$ with $i$ to create a smaller but similarly-shaped hexagon. This logic extends to other triangles as well; specifically, the truth values of $\holep{c}{d}{e}$ and $\holep{a}{e}{f}$ are not necessary to infer the presence of a $6$-hole.

Our insight emerged when we noticed that the SAT solver eliminated some $3$-hole literals from previous encodings. This elimination occurred primarily when only a few points existed between the leftmost and rightmost points of a triangle. On the other hand, the solver struggles significantly to identify the redundancy of these $3$-hole literals when the leftmost and rightmost points of a triangle were far apart. Therefore, to enhance the encoding's effectiveness, we chose to omit these $3$-hole literals (instead of letting the solver figure it out). 

Blocking the existence of a $6$-hole within the $6$-gon described above can be achieved with the following clause (which simply negates the assignment):
\begin{eqnarray}
\orientp{a}{b}{c} \lor \orientp{b}{c}{d} \lor \orientp{c}{d}{e} \lor \orientp{d}{e}{f}  \lor \holen{a}{c}{e}
\label{eq:explicit_holes}
\end{eqnarray}

For each set of six points, 16 different configurations can result in a $6$-hole. These configurations depend on which points are positioned above or below the line connecting the leftmost and rightmost points among the six. Three types of such configurations are illustrated in Fig.~\ref{fig:holes}, while the remaining configurations are symmetrical. It is important to note that this adds $16 \times \binom{n}{6}$ clauses to the formula, significantly increasing its size. However, in Section~\ref{sec:impact}, we will show that this improves performance.

We can reduce the number of clauses by about 30\% by strategically selecting which triangle within a $6$-gon is checked to be empty (i.e., which $3$-hole literal will be used). The two options are the triangle that includes the leftmost point (as depicted in Fig.~\ref{fig:holes}) and the triangle with the second-leftmost point. If the leftmost point is $p_1$, we opt for the second-leftmost point; otherwise, we choose the leftmost point. After propagating the unit clauses $\orientp{1}{a}{b}$, the clauses that describe configurations with three points below the line \aline{a}{\mathit{f}} are subsumed by the clause for the configuration with four points below the line \aline{1}{\mathit{f}}.

\subsection{An $O(n^4)$ Encoding}
\label{sec:aux}

This section is rather technical. It introduces auxiliary variables to reduce our encoding to $O(n^4)$ clauses. The process is known as structured bounded variable addition (SBVA)~\cite{SBVA}, which in each step adds a new auxiliary variable to encode a subset of the formula more compactly. SBVA heuristically selects the auxiliary variables. Instead, we select them manually because it is more effective, the new variables have meaning, and SBVA is extremely slow on this problem. Eliminating the auxiliary variables results in the encoding of Section~\ref{sec:arc}.

The first type of these variables, $\xfour_{a,c,d}$, represents the presence of a $4$-gon $\{a,b,c,d\}$
such that points $a,b,c,d$ appear in this order from left to right and $b$ and $c$ are above the line \aline{a}{d}. 
Furthermore, the variables $\xfive_{a,d,e}$ indicate the existence of a $5$-gon $\{a,b,c,d,e\}$ 
with the property that 
the points $a,b,c,d,e$ appear in this order from left to right,
the points $b$, $c$, and $d$ are above the line \aline{a}{e}, and 
the triangle $\{a,c,e\}$ is empty.
This configuration implies the existence of a $5$-hole within $\{a,b,c,d,e\}$ using similar reasoning as described in Section~\ref{sec:arc}. The logic enforcing these properties is outlined below. 
\begin{eqnarray}
\orientn{a}{b}{c} \land \orientn{b}{c}{d} \rightarrow \xfour_{a,c,d}&~~~~~& \mathrm{with~}a<b<c<d
\label{eq:xfour}\\
{\xfour_{a,c,d}} \land \orientn{c}{d}{e}  \land \holep{a}{c}{e} \rightarrow \xfive_{a,d,e} 
&~~~~~& \mathrm{with~}a<c<d<e
\label{eq:xfive}
\end{eqnarray}

In the following we distinguish five types of 6-holes 
by the number of points that lie above/below the line connecting the leftmost and rightmost points. 
Fig.~\ref{fig:holes} shows three configurations with four, three, and two points above the line, respectively. 
The configurations with three and four points below the line are symmetric 
but will be handled in a different and more efficient manner below.

To block all $6$-holes with configurations having three or four points above the line connecting the leftmost and rightmost points,
we utilize the variables $\xfive_{a,d,e}$.
Specifically, a configuration with three points above occurs if there is a point $b$ situated between $a$ and $e$, lying below the line \aline{a}{e}.  
Also, the configuration with four points above arises when a point $f$, located to the right of $e$, falls below the line \aline{d}{e}.
The associated clauses for these configurations are detailed below. 
The omission of 3-hole literals is justified by our knowledge that a $3$-hole exists among $a$, $c$, and $e$ for some point~$c$ positioned above the line \aline{a}{e}.
\begin{eqnarray}
\overline{\xfive_{a,d,e}} \lor \orientn{a}{b}{e} &~~~~~& \mathrm{with~}a < d <e, a < b < e 
\label{eq:threeabove}\\
\overline{\xfive_{a,d,e}} \lor \orientp{d}{e}{f}  &~~~~~& \mathrm{with~}a < d < e < f
\label{eq:fourabove}
\end{eqnarray}

To block the third type of 6-hole,
we need to introduce variables $\yfour_{a,c,d}$ which, 
similar as $\xfour_{a,c,d}$,
indicate the presence of a $4$-gon $\{a,b,c,d\}$
with the property that the points $a,b,c,d$ appear in this order from left to right and $b$ and $c$ are \emph{below} the line \aline{a}{d}. 
The logic that encode these variables is shown below.
\begin{eqnarray}
\orientp{a}{b}{c} \land \orientp{b}{c}{d} \rightarrow \yfour_{a,c,d} &~~~~~& \mathrm{with~}a<b<c<d
\label{eq:yfour}
\end{eqnarray}

Using the variables $\xfour_{a,c,d}$ and $\yfour_{a,c'\!,d}$ we are now ready to block the configuration of the third type of a 6-hole where two points lie above and two points lie below the line connecting the leftmost and rightmost points; see Fig.~\ref{fig:holes} (right). Recall that $\xfour_{a,c,d}$ denotes a $4$-gon situated above the line \aline{a}{d}, with $c$ being the second-rightmost point. Also, $\yfour_{a,c'\!,d}$ denotes a $4$-gon below the line \aline{a}{d}, with $c'$ as the second-rightmost point. A $6$-hole exists if both $\xfour_{a,c,d}$ and $\yfour_{a,c',d}$ are true for some points $a$ and $d$ when there are no points within the triangle formed by $a$, $c$, and $c'$. Or, in clauses:
\begin{eqnarray}
\overline{\xfour_{a,c,d}} \lor \overline{\yfour_{a,c'\!,d}} \lor \holen{a}{c}{c'} &~~~~~& \mathrm{with~}a<c<c'<d\\
\overline{\xfour_{a,c,d}} \lor \overline{\yfour_{a,c'\!,d}} \lor \holen{a}{c'}{c} &~~~~~& \mathrm{with~}a<c'<c<d
\label{eq:twotwo}
\end{eqnarray}

The remaining configurations to consider involve those with three or four points below the line joining the leftmost and rightmost points. As we discussed at the end of Section~\ref{sec:arc}, these configurations can be encoded more compactly. We only need to block the existence of $5$-holes $\{a,b,c,d,e\}$ with the property that the points $1,a,b,c,d,e$ appear in this order from left to right and
the points $b$, $c$, and $d$ are below the line \aline{a}{e}. 
The reasoning is as follows: if such a $5$-hole exists, it can be expanded into a $6$-hole by the closest point to line \aline{a}{b} within the triangle $\{1,a,b\}$. If the triangle is empty, this is point 1.
Additionally, by blocking these specific $5$-holes, we simultaneously block all $6$-holes with three or four points below the line between the leftmost and rightmost points. Following the earlier cases, we only require a single $3$-hole literal which ensures that the triangle $\{a,c,e\}$ is empty. The clauses to block these $5$-holes are as follows:
\begin{eqnarray}
\overline{\yfour_{a,c,d}} \lor \orientn{c}{d}{e} \lor \holen{a}{c}{e} &~~~~~& \mathrm{with~} 1 < a < c < d < e
\label{eq:threebelow}
\end{eqnarray}

This encoding uses $O(n^4)$ clauses, while it has the same propagation power as having all $16 \times \binom{n}{6}$ clauses in the domain-consistent encoding of Section~\ref{sec:arc}. 
In general, the trusted encoding
for $k$-holes uses $O(n^k)$ clauses, while the optimized encoding when generalized to $k$-holes has only $O(kn^4)$ clauses, or $O(n^4)$ for every fixed~$k$. 
An encoding of size $O(n^4)$ for $k$-gons is analogous: 
simply remove the $3$-hole literals from the clauses. 

\subsection{Minor Optimizations}
\label{sec:minor}

We can make the encoding even more compact by removing a large fraction of the clauses from the trusted encoding. 
Note that constraints to forbid $6$-holes contain only negative $3$-hole literals. 
That means that only half of the constraints to define the $3$-hole variables are actually required. 
This in turn shows that only half of the inside variable definitions are required. So, instead of (\ref{eq:inside1}), 
(\ref{eq:inside2}), and~(\ref{eq:hole}), 
it suffices to use the following:
\begin{eqnarray}
\insidep{i}{a}{b}{c} &\rightarrow& \Big(\big(\orientp{a}{b}{c} \rightarrow (\orientn{a}{i}{b} \land \orientp{a}{i}{c})\big) \land \big(\orientn{a}{b}{c} \rightarrow (\orientp{a}{i}{b} \land \orientn{a}{i}{c})\big)\Big)\\
\label{eq:inside1b}
\insidep{i}{a}{b}{c} &\rightarrow& \Big(\big(\orientp{a}{b}{c} \rightarrow (\orientp{a}{i}{c} \land \orientn{b}{i}{c})\big) \land \big(\orientn{a}{b}{c} \rightarrow (\orientn{a}{i}{c} \land \orientp{b}{i}{c})\big)\Big)\\
\label{eq:inside2b}
\holep{a}{b}{c} &\leftarrow& 
\bigwedge_{\substack{a<i<c\\ i \neq b}}
\insiden{i}{a}{b}{c}.
\label{eq:holeb}
\end{eqnarray}

It is worth noting that the SAT preprocessing technique blocked-clause elimination (BCE) 
will automatically remove the clauses we omit \cite{BCE}. 
However, for means of efficiency, BCE is turned off by default in top-tier solvers, including the solver {\sf CaDiCaL}, which we used for the proof. During initial experiments, we observed that omitting these clauses slightly improves the performance.

Finally, the variables $\xfour_{a,c,d}$ and $\yfour_{a,c,d}$ can be used to more compactly encode the clauses~(\ref{eq:axiom2}). We can replace the clauses (\ref{eq:axiom2}) with:
\begin{eqnarray}
(\overline{\xfour_{a,c,d}} \lor \orientn{a}{c}{d}) \land (\overline{\yfour_{a,c,d}} \lor \orientp{a}{c}{d})&~~~~~& \mathrm{with~}a<c<d
\label{eq:auxaxiom2}
\end{eqnarray}

\subsection{Breaking the Reflection Symmetry}
\label{sec:symmetry}

Holes are invariant to reflectional symmetry:
If we mirror a point set~$S$,
then the counterclockwise order around the extremal point~$p_1$ (which is $p_2,\ldots,p_n$) is reversed
(to $p_n,\ldots,p_2$).
By relabeling points to preserve the counterclockwise order, 
we preserve $\orientp{1}{a}{b}=true$ for $a<b$, while
the original orientation variables $\orientp{a}{b}{c}$ with $2 \le a<b<c \le n$ are mapped
to $\orientp{n-c+2}{n-b+2}{n-a+2}$.
A similar mapping applies to the containment and $3$-hole variables. 
The trusted encoding maps almost onto itself, except for the missing reflection clauses of (\ref{eq:axiom1}) and (\ref{eq:axiom2}). As a fix for  verification, we add each reflected clause using one resolution step. 

Since only a tiny fraction of triple orientations map to themselves (so-called \emph{involutions}),
breaking the reflectional symmetry reduces the search space by a factor of almost~2.
We partially break this symmetry by constraining the variables $\orientp{a}{a+1}{a+2}$ with $2 \leq a \leq n-2$.
We used the symmetry-breaking predicate below, because it is compatible with 
our cube generation, described in Section~\ref{sec:partitioning}.
\begin{equation}
\orientp{\lceil\frac{n}{2}\rceil-1}{\lceil\frac{n}{2}\rceil}{\lceil\frac{n}{2}\rceil+1},
\dots, 
\orientp{2}{3}{4}
\preccurlyeq 
\orientp{\lfloor\frac{n}{2}\rfloor+1}{\lfloor\frac{n}{2}\rfloor+2}{\lfloor\frac{n}{2}\rfloor+3},
\dots, 
\orientp{n-2}{n-1}{n}
\label{eq:sbp}
\end{equation}

One symmetry that remains is the choice of the first point. Any point on the convex hull could be picked for this purpose, and breaking it can potentially reduce the search space by at least a factor of 3. However, breaking this symmetry effectively is complicated, and we therefore left it on the table.

\section{Problem Partitioning} 
\label{sec:partitioning}

The formula to determine that $h(6) \leq 30$ requires CPU years to solve. To compute this in reasonable time, the problem needs to be partitioned into many small subproblems that can be solved in parallel. Although tools exist to construct partitionings automatically~\cite{HKWB2012_cubes}, we observed that this partitioning was ineffective. As a consequence, we focused on manual partitioning. 

During our initial experiments, we determined which orientation variables were suitable for splitting. We used the formula for $g(6) \leq 17$ for this purpose because its runtime is large enough to make meaningful observations and small enough to explore many options. It turned out that the orientation variables $\orientp{a}{a+1}{a+2}$ were the most effective choice for splitting the problem. Assigning one of these $\orientp{a}{a+1}{a+2}$ variables to true/false roughly halves the search space and reduces the runtime by a factor of roughly~2. 

A problem with $n$ points has $n-3$ free variables of the form $\orientp{a}{a+1}{a+2}$, as the variable $\orientp{1}{2}{3}$ is already fixed by the symmetry breaking.
One cannot generate $2^{n-3}$ equally easy subproblems, because $(\orientn{a}{a+1}{a+2} \lor \orientn{a+1}{a+2}{a+3} \lor \orientn{a+2}{a+3}{a+4})$ and $(\orientp{a}{a+1}{a+2} \lor \orientp{a+1}{a+2}{a+3} \lor \orientp{a+2}{a+3}{a+4} \lor \orientp{a+3}{a+4}{a+5})$ follow directly from the  optimized formula after unit propagation. Thus, assigning three consecutive $\orientp{a}{a+1}{a+2}$ variables to true results directly in a falsified clause, as it would create a 6-hole among the points $p_1$, $p_a$, $\dots$, $p_{a+4}$. The same holds for four consecutive $\orientp{a}{a+1}{a+2}$ variables assigned to false, which would create a 6-hole among the points $p_a$, $\dots$, $p_{a+5}$. The asymmetry is due to fixing the variables $\orientp{1}{a}{b}$ to true. If we assigned them to false, then the opposite would happen.

We observed that limiting the partition to variables involving the middle points reduces the total runtime. 
We will demonstrate such experiments in Section~\ref{sec:sixseven}.
So, to obtain suitable cubes,
we considered all assignments of the sequence 
$\orientp{a}{a+1}{a+2}$, $\orientp{a+1}{a+2}{a+3}$, $\ldots$, $\orientp{a+\ell-1}{a+\ell}{a+\ell+1}$ 
for a suitable constant $\ell$ and $a = \frac{n+\ell}{2} - 1$
such that the above properties are fulfilled, that is,
no three consecutive entries are true
and no four consecutive entries are false.
In the following we refer to $\ell$ as the \emph{length} of the cube-space.
In our experiments of Section~\ref{sec:impact}, we observed that picking $\ell < n - 3$
reduces the overall computational costs. 
Specifically, for the $h(6) \leq 30$ experiments, we use length $\ell = 21$.

Our initial experiments showed that the runtime of cubes grows exponentially with the number of occurrences of
the alternating pattern $\orientp{b}{b+1}{b+2} = +$, 
$\orientp{b+1}{b+2}{b+3} = -$, $\orientp{b+2}{b+3}{b+4} = +$. As a consequence,
the hardest cube for $h(6) \leq 30$ would still require days of computing time, thereby limiting parallelism. 
To deal with this issue, we further partition cubes that contain this pattern. For each occurrence of the alternating pattern in a cube,
we split the cube into two cubes: one that extends it with $\orientp{b}{b+2}{b+4}$
and one that extends it with $\orientn{b}{b+2}{b+4}$. Note that we do this for each occurrence. So a cube containing $m$ of these patterns is split into $2^m$ cubes. This reduced the computational costs of the hardest cubes to less than an hour.

\goodbreak

\section{Evaluation}

For the experiments, we use the solver {\sf CaDiCaL} (version 1.9.3)~\cite{BFFH2020}, which is currently the only top-tier solver that can produce LRAT proofs directly. The efficient, verified checker {\sf cakeLPR} \cite{cakelprsttt} validated the proofs.  
We run {\sf CaDiCaL} with command-line options: {\tt ----sat} {\tt ----reducetarget=10} {\tt ----forcephase} {\tt ----phase=0}. The first option reduces the number of restarts. This is typically more useful for satisfiable formulas (as the name suggests), but in this case it is also helpful for unsatisfiable formulas. The second option turns off aggressive clause deletion strategy, which is usually helpful for large formulas. The last two options tell the solver to assign decision variables to false, a {\sf MiniSAT} heuristic~\cite{MiniSAT}. Each of these settings improved performance compared to the default setting on the formulas used in the evaluation. 
Experiments were run on a specialized, internal Amazon Web Services solver framework that provides cloud-level scaling. The framework used {\tt m6i.xlarge} instances, which have two physical cores and 16 GB of memory.

\subsection{Impact of the Encoding}
\label{sec:impact}

To illustrate the impact of the encoding on the performance, we show some statistics on various encodings of the $h(6) \leq 30$ formula. We restricted this experiment to solving a single randomly-picked subproblem. For other subproblems, the results were similar. We experimented with five encodings:
\begin{itemize}
\item $T$: the trusted encoding presented in Section~\ref{sec:encoding}
\item $O_1$: $T$ with (\ref{eq:holes_constraints}) replaced by the domain-consistent encoding (\ref{eq:explicit_holes}) of Section~\ref{sec:arc} 
\item $O_2$: $O_1$ with (\ref{eq:explicit_holes}) replaced by the $O(n^4)$ encoding of Section~\ref{sec:aux}
\item $O_3$: $O_2$ with the minor optimizations that replace  (\ref{eq:inside1}), (\ref{eq:inside2}), (\ref{eq:hole}), and (\ref{eq:axiom2}) by (\ref{eq:inside1b}), (\ref{eq:inside2b}), (\ref{eq:holeb}), and (\ref{eq:auxaxiom2}), respectively, see Section~\ref{sec:minor}
\item $O_4$: $O_3$ extended with the symmetry-breaking predicate from Section~\ref{sec:symmetry}
\end{itemize}

\begin{table}[b]
\caption{Comparison of the different encodings of randomly-picked subproblem}
\label{tab:encoding}
\centering
\begin{tabular}{c@{~~~}r@{~~~}r@{~~~}r@{~~~}r@{~~~}r@{~~~}r@{~~~}c}
\toprule
formula & $\#$variables & $\#$clauses & $\#$conflicts & $\#$propagations & time (s) \\
\midrule
$T$     & 62\,930~~ & 1\,171\,942 & 1\,082\,569 & 1\,338\,662\,627 & 243.07\\
$O_1$ & 62\,930~~ & 5\,823\,078 &    228\,838 &     282\,774\,472 & 136.20\\
$O_2$ & 75\,110~~ &     667\,005 &    211\,272 &      343\,388\,591 &  45.49\\
$O_3$ & 75\,110~~ &     436\,047 &    234\,755 &      340\,387\,692 &  39.46\\
$O_4$ & 75\,110~~ &     444\,238 &    234\,587 &      342\,904\,580 &  39.41\\
\bottomrule
\end{tabular}
\end{table}

Table~\ref{tab:encoding} summarizes the results. The domain-consistent encoding can be solved more efficiently than the trusted encoding while having over five times as many clauses. The reason for the faster performance becomes clear when looking at the number of conflicts and propagations. The domain-consistent encoding requires just over a fifth as many conflicts and propagations to determine unsatisfiability. The auxiliary variables that enable the $O(n^4)$ encoding reduce the size by almost an order of magnitude. The resulting formula can be solved three times as fast, while using a similar number of conflicts and propagations. The minor optimizations reduce the size by roughly a third and further improve the runtime. Finally, the addition of the symmetry-breaking predicate doesn't impact the performance. Its main purpose is to halve the number of cubes.

We also solved the optimized encoding ($O_3$) of the formula $g(6) \leq 17$, which takes 41.99 seconds using 623\,540 conflicts. Adding the symmetry-breaking predicate ($O_4$) reduces the runtime to 17.39 seconds using 316\,785 conflicts. So the symmetry-breaking predicate reduces the number of conflicts by roughly a factor of 2 (as expected) while the runtime is reduced even more. The latter is due to the slowdown caused by maintaining more conflict clauses while solving the formula without the symmetry-breaking predicate.

\begin{table}[b]
\caption{Runtime comparison for Theorem~\ref{thm:h6g7=24} using different values of parameter $\ell$}
\label{tab:partitions}
\centering
\begin{tabular}{@{~~~}r@{~~~}r@{~~~}r@{~~~}r@{~~~}r@{~~~}r@{~~~}c}
\toprule
 $\ell$~ & $\#$cubes & average time (s) & max time (s) & total time (h)\\
\midrule
21 & 312\,418 & 6.99~~~~  & 66.86~~~ & 606.55~~~\\
19 & 89\,384 & 13.61~~~~ & 123.70~~~ & 337.96~~~\\
17 & 25\,663 & 34.29~~~~ & 293.10~~~ & 244.50~~~\\
15 & 7393 & 112.61~~~~ & 949.50~~~ & 231.27~~~\\
13 & 2149 & 431.26~~~~ & 3\,347.59~~~ & 257.44~~~\\
11 & 629 & 1\,847.46~~~~ & 11\,844.05~~~ & 322.79~~~\\
9 & 188 & 7\,745.14~~~~ & 32\,329.05~~~ & 404.47~~~\\
7 & 57 & 32\,905.90~~~~ & 105\,937.76~~~ & 521.01~~~\\
\bottomrule
\end{tabular}
\vspace{-10pt}
\end{table}

\subsection{Impact of the Partitioning}
\label{sec:sixseven}

\begin{figure}[t]
    \centering
    \input{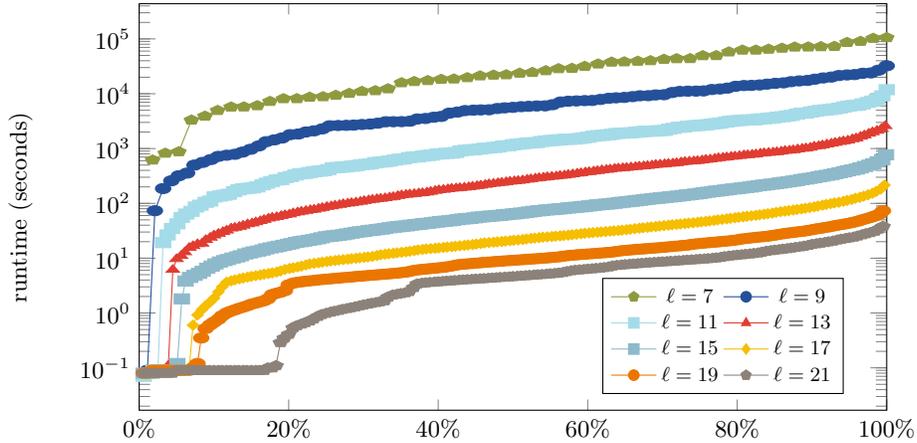}
    \vspace{-15pt}
    \caption{Runtime to solve the subproblems of Theorem~\ref{thm:h6g7=24} for various splitting parameters}
    \label{fig:evaluation}
\end{figure}

All known point sets witnessing the lower bound $h(6) \ge 30$ contain a $7$-gon. To obtain a possibly easier problem to test and compare heuristics, 
we studied how many points are required to guarantee the existence of a $6$-hole or a $7$-gon. It turned out that the answer is at most~24 (Theorem~\ref{thm:h6g7=24}). 
Computing this is still hard but substantially easier compared to our main result. During our experiments, we 
observed that increasing the number of cubes eventually increase the total runtime. We therefore explored which parameters
produce the lowest total runtime. The experimental results are shown in Table~\ref{tab:partitions} for various values
for the parameter $\ell$. Incrementing $\ell$ by 2 increases the number of cubes roughly by a factor of 3.
The optimal total runtime is achieved for $\ell =15$, which is a 62\% reduction compared to full
partitioning ($\ell = 21$). Note that the solving time for the hardest cube (the max column) increases 
substantially when using fewer cubes. This in turn reduces the effectiveness of parallelism. The runtime without partitioning is expected to be about 1000 CPU hours,
so partitioning achieves super-linear speedups and more than a factor of 4 speedup for $\ell=15$. 
Fig.~\ref{fig:evaluation} shows plots of cumulatively solved cubes, with similar curves for all settings. 

We also evaluated the off-the-shelf tool {\sf March} for partitioning. This tool was used to prove Schur Number Five~\cite{Schur}. We used option {\tt -d 13} to cut off partitioning at depth 13 to create 8192 cubes. That partition turned out to be very poor: at least 18 cubes took over 100\,000 seconds. The expected total costs are about 10\,000 CPU hours, so 10 times the estimated partition-free runtime. 

A partitioning can also guide the search to solve the formula $g(6) \leq 17$. The partitioning of this formula using $\ell = 12$ results in 1108 cubes. If we add these cubes to the formula with the symmetry-predicate ($O_4$) in the iCNF format~\cite{iCNF}, then CaDiCaL can solve it in 8.53 seconds using 205\,153 conflicts.

\subsection{Theorem~\ref{thm:h6=30}}

\begin{figure}[t]
    \centering
    \input{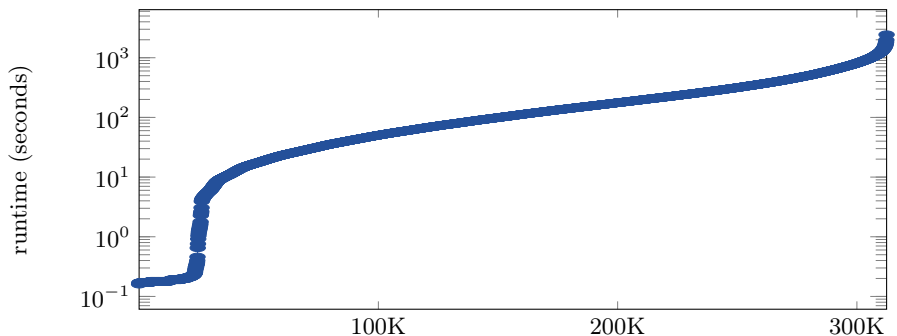}
    \caption{Reported process time to solve the subproblems of $h(6) \leq 30$ with proof logging 
    while running the cakeLPR verified checker on another core.}
    \label{fig:theorem1}
\end{figure}

To show that the optimized encoding for $h(6) \leq 30$ is unsatisfiable, we partitioned the Theorem~\ref{thm:h6=30} problem with the splitting algorithm described in Section~\ref{sec:partitioning} with parameter $\ell = 21$, which results in $312\,418$ cubes. 
We picked this setting based on the experiments shown in Table~\ref{tab:partitions}. Fig.~\ref{fig:theorem1}
shows the runtime of solving the subproblems. The average runtime was just below 200 seconds. All subproblems were solved in less than an hour. 
Almost $24\,000$ subproblems could be solved within a second. For these subproblems, the cube resulted directly in a conflict, 
so the solver didn't have to perform any search.

The total runtime is close to 17\,300 CPU hours, 
or slightly less than 2 CPU years. We could achieve practically a linear speedup using 1000 {\tt m6i.xlarge}
instances. The timings include producing and validating the LRAT proof. We chose the LRAT proof format, because it allows concurrent checking, as described in Section~\ref{sec:concurrent}. The combined size of the proofs is 180 terabytes in the uncompressed LRAT format used by the cakeLPR  checker. In past verification efforts of hard math problems, the produced proofs were in the DRAT format. For this problem, the LRAT proofs are roughly 2.3 times as large as the corresponding DRAT proof.
We estimate that the DRAT proof would have been 78 terabytes in size,
so approximately one third  of the Pythagorean Triples proof~\cite{ptn}.
For all problems, the checker was able to easily keep up with the solver while running on a different core, thereby finishing as soon as the solver was done. 

\subsection{Lower-Bound Experiments}
\label{sec:overmars}

\begin{figure}[t]
    \begin{minipage}{\textwidth}
\begin{tikzpicture}[scale=0.9]

  \node at (8,12.5) {
  coordinates:};
  
  \node at (8,10) {
\begin{tabular}{r@{~~\,}r}
1 &1260 \\
16  &743\\
22 & 531\\
37 & 0\\
306 & 592\\
310 & 531\\
366 & 552\\
371 & 487\\
374 & 525\\
~& ~\\
  \end{tabular}
  };

\node at (10,10) {
\begin{tabular}{r@{~~\,}r}
392 & 575 \\
396 & 613 \\
410 & 539 \\
416 & 550 \\
426 & 526 \\
434 & 552 \\
436 & 535 \\
446 & 565 \\
449 & 518 \\
450 & 498 \\
  \end{tabular}
  };  
  
\node at (12,10) {
\begin{tabular}{r@{~~\,}r}
453 & 542 \\
458 & 526 \\
489 & 537 \\
492 & 502 \\
496 & 579 \\
516 & 467 \\
552 & 502 \\
754 & 697 \\
777 & 194 \\
1259 & 320 \\
  \end{tabular}
  };  
  
  \tikzset{vertex/.style={draw, circle, fill=blue, scale=0.3}}
     
\node[vertex] (aa) at (-0.05,12.60) {};
\node[vertex] (ab) at (.31,0) {};
\node[vertex] (ac) at (12.85,3.16) {};

\node[vertex] (bb) at (.16,7.43) {};
\node[vertex] (ba) at (.22,5.31) {};
\node[vertex] (c) at (3.06,5.92) {};
\node[vertex] (d) at (3.10,5.31) {};
\node[vertex] (e) at (3.66,5.52) {};
\node[vertex] (f) at (3.71,4.87) {};
\node[vertex] (g) at (3.74,5.25) {};
\node[vertex] (h) at (3.92,5.75) {};
\node[vertex] (i) at (3.96,6.13) {};
\node[vertex] (j) at (4.10,5.39) {};
\node[vertex] (k) at (4.16,5.50) {};
\node[vertex] (l) at (4.26,5.26) {};
\node[vertex] (m) at (4.34,5.52) {};
\node[vertex] (n) at (4.36,5.35) {};
\node[vertex] (o) at (4.46,5.65) {};
\node[vertex] (p) at (4.49,5.18) {};
\node[vertex] (q) at (4.50,4.98) {};
\node[vertex] (r) at (4.53,5.42) {};
\node[vertex] (s) at (4.58,5.26) {};
\node[vertex] (t) at (4.89,5.37) {};
\node[vertex] (u) at (4.92,5.02) {};
\node[vertex] (v) at (4.96,5.79) {};
\node[vertex] (w) at (5.16,4.67) {};
\node[vertex] (x) at (5.52,5.02) {};
\node[vertex] (bc) at (7.54,6.97) {};
\node[vertex] (bd) at (7.77,1.94) {};

\draw[] (aa) -- (ab) -- (ac) -- (aa);
\draw[] (ba) -- (bb) -- (bc) -- (bd) -- (ba);

\draw[] (x) -- (v) -- (i) -- (c) -- (d) -- (f) --(w) -- (x)  ;

\draw[] (t) -- (o)-- (h) -- (e) -- (g) --(q) --(u) --(t)  ;

\draw[] (s) -- (r) -- (m)-- (k) -- (j) -- (l) -- (p) -- (s)  ;

\end{tikzpicture}

\end{minipage}
	\caption{A set of 29 points with no $6$-hole and no $8$-gon~\cite{Overmars2002}.
	The three points forming the convex hull are slightly moved outward to avoid the visual confusion that some points appear collinear.
	The lines show the six convex hull layers.}
	\label{fig:overmars}
\end{figure}
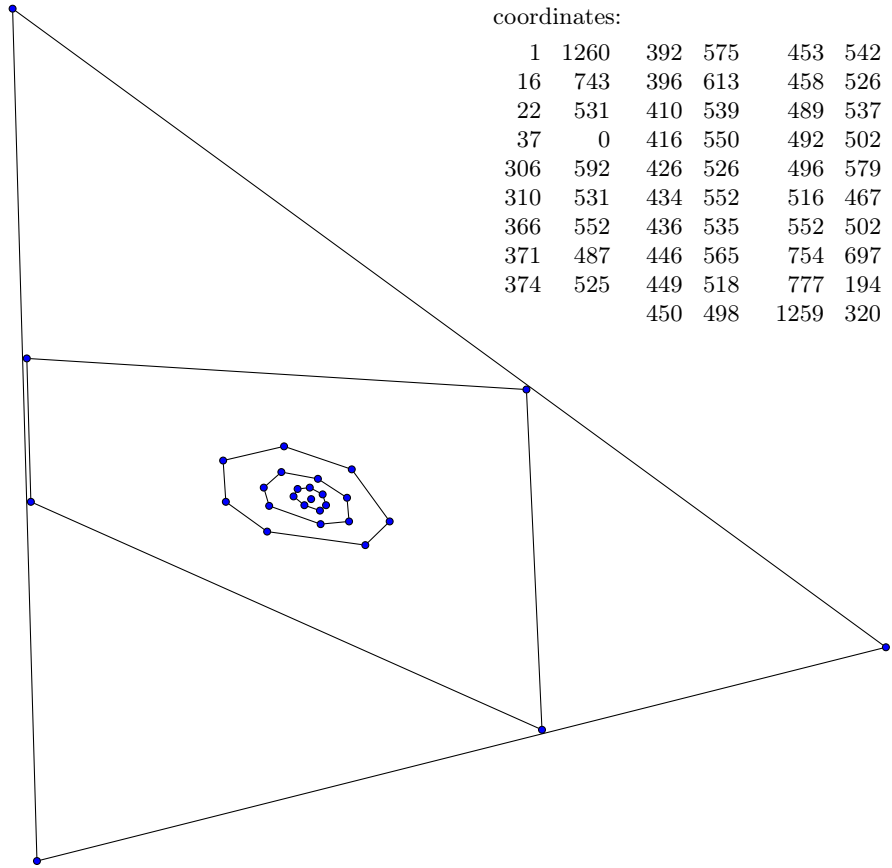

Overmars constructed a 29-point set without $6$-hole~\cite{Overmars2002}, see Fig.~\ref{fig:overmars}. 
The layers of the convex hull have size 3, 4, 7, 7, 7, 1. The paper mentioned
that the convex hull layers of all $6$-hole-free 29-point set found by the local search were the same. 

We used our encoding to find many $6$-hole-free 29-point sets. We partitioned the problem using $\ell = 22$, which 
results in $581\,428$ cubes. Out of those cubes, $116\,305$ ($20.00\%$) were satisfiable. For all the cubes, the first 
solution found by the solver had the same layers of the convex hull. We also tested for each of these cubes whether
there is a solution for which either the first layer has more than 3 points or the second layer has exactly three points.
This can be done by adding a single clause to the formula asking whether there is a point below the line \aline{p_{2}}{p_{29}}
or whether point $p_{4}$ is in the triangle $\{p_{3},p_{27},p_{28}\}$ or $p_{27}$ is in the triangle $\{p_{3},p_{27},p_{28}\}$.
Adding that clause made all cubes unsatisfiable.

The result above means that all $6$-hole-free 29-point sets have exactly 3 points in the convex hull and the next layer has
at least 4 points. Note that this implies that there cannot be a $6$-hole-free 30-point set.

Although we haven't verified it yet, it seems likely that the convex hull layers of  all $6$-hole-free 29-point sets are the same. 
As a consequence, each of those point sets has at least three $7$-gons. 

\section{Verification}
\label{sec:verification}

We applied three verification steps to increase trust in the correctness of our results. In the first step, we check the results produced by the SAT solver. The second step consists of checking the correctness of the optimizations discussed in Section~\ref{sec:optimization}.
In the third step, we validate that the case split covers all cases. 


\vspace{-10pt}

\subsection{Concurrent Solving and Checking}
\label{sec:concurrent}

The most commonly used approach to validate SAT-solving results works as follows. First, a SAT solver produces a DRAT proof. This proof is checked and trimmed using a fast, but unverified tool that produces a LRAT proof. The difference between a DRAT proof and a LRAT proof is that the latter contains hints. The LRAT proof is then validated by a formally-verified checker, which uses the hints to obtain efficient performance.

Recently, the SAT solver {\sf CaDiCaL} added support for producing LRAT proofs directly (since version 1.7.0). This allows us to produce the proof and validate it concurrently. To the best of our knowledge, we are the first to take advantage of this possibility. {\sf CaDiCaL} sends its proof to a unix pipe and the verified checker {\sf cakeLPR} reads it from the pipe. 
This tool chain works remarkably well, adds little performance overhead, and avoids needing to store large files.

\vspace{-10pt}

\subsection{Reencoding Proof}
We validated the four optimizations presented in Section~\ref{sec:optimization}. Only the trusted encoding has the reflection symmetry, as none of the optimizations preserve this symmetry. Each of the clauses in the symmetry-breaking predicate have the substitution redundancy (SR) property~\cite{BussThapen2021} with respect to the trusted encoding. However, there doesn't exist a SR checker. Instead, we transformed the SR check into a sequence of DRAT addition and deletion steps. This is feasible for small point sets (up to 10), but is too expensive for the full problem. It may therefore be more practical to verify this optimization in a theorem prover. 

Transforming the trusted encoding into the domain-consistent one is challenging to validate because the solver cannot easily infer the existence of a $6$-hole using only the clauses (\ref{eq:explicit_holes}). Since we are replacing (\ref{eq:holes_constraints}) by (\ref{eq:explicit_holes}) and clause deletion trivially preserves satisfiability, we only need to check whether each of the clauses (\ref{eq:explicit_holes}) is entailed by the trusted encoding. This can be achieved by constructing a formula that asks whether there exists an assignment that satisfies the trusted encoding, but falsifies at least one of the clauses (\ref{eq:explicit_holes}). We validated that this formula is unsatisfiable for $n \leq 12$ (around 300 seconds).\footnote{We implemented an entailment tool, see \url{https://github.com/marijnheule/entailment}} The formula becomes challenging to solve for larger $n$. However, the validation for small $n$ provides substantial evidence of the correctness of the encoding and the implementation.  

Checking the correctness of the other two optimizations is easier. Observe that one can obtain the domain-consistent encoding from the $O(n^4)$ encoding by applying Davis-Putnam resolution~\cite{DP60} on the auxiliary variables. This can be expressed using DRAT steps. The DRAT derivation from the domain-consistent encoding to the $O(n^4)$ encoding applies all these steps in reverse order. The minor optimizations mostly delete clauses, which is trivially correct for proofs of unsatisfiability. The clauses (\ref{eq:auxaxiom2}) have the RAT property on the auxiliary variables and their redundancy is easily checked using a DRAT checker. 

\vspace{-10pt}

\subsection{Tautology Proof}

The final validation step consists of checking whether the partition of the problem covers the entire search space. This part has also been called the tautology proof~\cite{Schur}, because in most cases it needs to determine whether the disjunction of cubes is a tautology. We take a slightly different approach and validate that the following formula is unsatisfiable: the conjunction of the negated cubes; the symmetry-breaking predicate; and some clauses from the formula. 

Recall that we omitted various cubes because they resulted in a conflict with the clauses $(\orientn{a}{a+1}{a+2} \lor \orientn{a+1}{a+2}{a+3} \lor \orientn{a+2}{a+3}{a+4})$ with $a \in \{2, \dots, n-4\}$ and $(\orientp{a}{a+1}{a+2} \lor \orientp{a+1}{a+2}{a+3} \lor \orientp{a+2}{a+3}{a+4} \lor \orientp{a+3}{a+4}{a+5})$ with $a \in \{2, \dots, n-5\}$. We checked with DRATtrim that these clauses are implied by the optimized formulas, which takes 0.3 CPU seconds in total. We combined them with the negated cubes and the symmetry-breaking predicate, which results in an unsatisfiable formula that can be solved by {\sf CaDiCaL} in 12 CPU seconds.

\section{Conclusion}

We closed the final case regarding $k$-holes in the plane by showing $h(6) = 30$. This is another example
that SAT-solving techniques can effectively solve a range of long-standing open problems in mathematics. Other successes include the Pythagorean Triples problem~\cite{ptn}, Schur Number Five~\cite{Schur}, and Keller's Conjecture~\cite{Keller}. Also, we recomputed $g(6) = 17$ many orders of magnitude faster compared to the original computation by Szekeres and Peters~\cite{SzekeresPeters2006} even when taking into account the difference in hardware. SAT techniques overwhelmingly outperformed their dedicated approach. Key contributions include an effective, compact encoding and a partitioning strategy enabling linear-time speedups even when using thousands of cores. We also presented a new concurrent proof-checking procedure to significantly decrease proof verification costs.

Although the tools are fully automatic, several aspects of our solution require significant user ingenuity.
In particular, we had to develop encoding optimizations and a search-space partitioning strategy to fully leverage the power of the tools. Constructing the domain-consistent encoding automatically appears challenging. Most other optimizations can be achieved automatically, for example via structured bounded variable elimination~\cite{SBVA}. However, the resulting formula cannot be solved nearly as efficiently as the presented one. Substantial research into generating effective partitionings is required to enable non-experts to solve such problems.
Although we validated most optimization steps, formally verifying the trusted encoding or even the domain-consistent encoding would further increase trust in the correctness of our result.  

\subsubsection{Acknowledgements} 

    Heule is partially supported by NSF grant CCF-2108521.
    Scheucher was supported by the DFG grant SCHE~2214/1-1.
    We thank Donald Knuth, Benjamin Kiesl-Reiter, John Mackey, Robert Jones, and the reviewers for their valuable feedback. 
    The authors met for the first time during Dagstuhl Seminar 23261 \enquote{SAT Encodings and Beyond}, 
    which kicked off the research published in this paper. 
    \iftacas
    \else
    We thank Helena Bergold for the visualization in Fig.~\ref{fig:n23g7h6_wiring}.
    \fi


\bibliographystyle{splncs04}
\bibliography{bibliography}

\iftacas
\else
\appendix

\section{Proof of Lemma~\ref{lemma:increasing_coordinates}}
\label{app:proof_of_lemma}

In the following proof, which is based on \cite{Scheucher2020}, 
we utilize the fact that, 
the triple orientation
$\orientp{a}{b}{c} = true$ encodes whether 
the sign of the determinant
\[
\det \begin{pmatrix}
	1  & 1  & 1  \\
	x_a& x_b& x_c\\
	y_a& y_b& y_c
\end{pmatrix} 
\] is positive,
and use
some basics from linear algebra.

\begin{proof}
First, we apply an affine-linear transformation to~$S$ 
so that $p_1$ is mapped to the origin $(0,0)$ and all other $p_i$, $i \ge 2$, have positive $x$- and $y$-coordinates.
To see this, apply a translation $(x,y) \mapsto (x+s,y+t)$ for some constants $s,t \in \mathbb{R}$
so that $p_1$ is mapped to the origin.
Since $p_1$ is an extremal point,
we can perform a rotation $(x,y) \to (x \cos(\phi)- y \sin(\phi),x \sin(\phi)+ y \cos(\phi))$  for some constant $\phi \in [0,2\pi)$ such that all points $p_2,\ldots,p_n$ have positive $x$-coordinate.
Finally, we apply a shearing transformation  $(x,y) \mapsto (x,y+c \cdot x)$ for some constant $c \in \mathbb{R}$ so that $p_2,\ldots,p_n$ have positive $y$-coordinate as well.
Pause to note that affine-linear transformations do not affect determinants and hence the triple orientations are persevered.
Formally, one can introduce transformation matrices to write the translation as
\[
\begin{pmatrix}
1 \\
x+s \\
y+t \\
\end{pmatrix} 
=
\begin{pmatrix}
1 & 0 & 0\\
s & 1 & 0\\
t & 0 & 1\\
\end{pmatrix} 
\cdot 
\begin{pmatrix}
1 \\
x \\
y \\
\end{pmatrix},
\]
a shearing as
\[
\begin{pmatrix}
1 \\
x \\
y+cx \\
\end{pmatrix} 
=
\begin{pmatrix}
1 & 0 & 0\\
0 & 1 & 0\\
0 & c & 1\\
\end{pmatrix} 
\cdot 
\begin{pmatrix}
1 \\
x \\
y \\
\end{pmatrix},
\]
and a rotation as
\[
\begin{pmatrix}
1 \\
x \cos(\phi)- y \sin(\phi) \\
x \sin(\phi)+ y \cos(\phi) \\
\end{pmatrix} 
=
\begin{pmatrix}
1 & 0 & 0\\
0 & \cos(\phi)  &  -\sin(\phi)\\
0 & \sin(\phi) & \cos(\phi))\\
\end{pmatrix} 
\cdot 
\begin{pmatrix}
1\\
x \\
y \\
\end{pmatrix}.
\]
Since each of the transformation-matrices has determinant~1,
and 
\[
\det \left( A \cdot 
\begin{pmatrix}
1 & 1 & 1\\
x_a & x_b & x_c\\
y_a & y_b & y_c\\
\end{pmatrix} \right) = \det(A ) \cdot \det \begin{pmatrix}
1 & 1 & 1\\
x_a & x_b & x_c\\
y_a & y_b & y_c\\
\end{pmatrix},
\]
none of these affine transformations affects the triple orientations.

Now
$x_i/y_i$ is increasing for $i \ge 2$ as  $p_2,\ldots,p_n$ are sorted counterclockwise around~$p_1$.
Since $S$ is in general position, 
there is an $\varepsilon > 0$ such that
$S$ and $S' := \{(0,\varepsilon)\} \cup \{p_2,\ldots,p_n\}$ are of the same order type.  Formally,
since the determinant is a polynomial and hence continuous,
it holds
\[
\sgn \det
\begin{pmatrix}
1 & 1 & 1\\
0 & x_a & x_b\\
\varepsilon & y_a & y_b\\
\end{pmatrix} 
=
\sgn \det
\begin{pmatrix}
1 & 1 & 1\\
0 & x_a & x_b\\
0 & y_a & y_b\\
\end{pmatrix} 
\]
for some sufficiently small $\varepsilon > 0$.
We next apply the projective transformation
\mbox{$(x,y) \mapsto (\nicefrac{x}{y},\nicefrac{-1}{y})$}
to $S'$ to obtain~$\tilde{S}$. 
By the multilinearity of the determinant, we obtain
\[
\det 
\begin{pmatrix}
1 & 1 & 1\\
x_a & x_b & x_c\\
y_a & y_b & y_c\\
\end{pmatrix}
=
y_a \cdot y_b \cdot y_c \cdot 
\det 
\begin{pmatrix}
1 & 1 & 1\\
 \nicefrac{x_a}{y_a} &  \nicefrac{x_b}{y_b} &  \nicefrac{x_c}{y_c}\\
\nicefrac{-1}{y_a} & \nicefrac{-1}{y_b} & \nicefrac{-1}{y_c}\\
\end{pmatrix}.
\]
Since all points in $S'$ have positive $y$-coordinates,
the signs of the determinants coincide, and hence
$S'$ and $\tilde{S}$ have the same triple orientations.
Moreover, as $\tilde{x_i}=\nicefrac{x_i'}{y_i'}$ is increasing for $i \ge 1$, 
the set $\tilde{S}$ fulfills all desired properties.
\qed
\end{proof}

\section{Realizability}
\label{sec:discussion}

\begin{figure}[t]
    \centering
	\includegraphics[width=\textwidth]{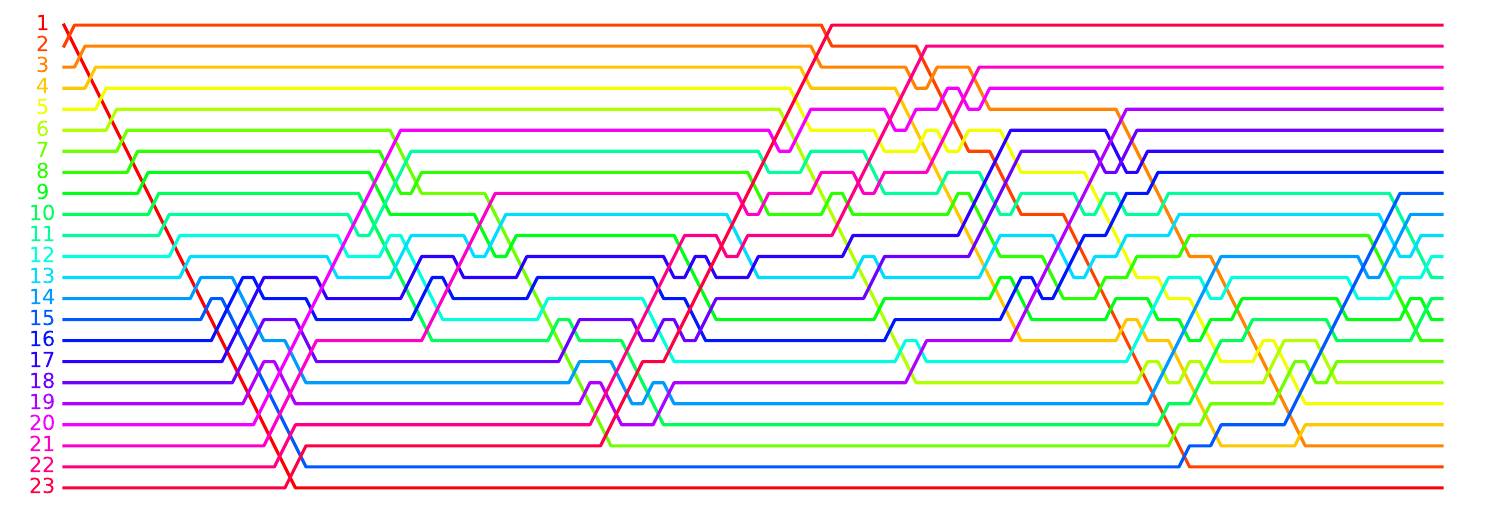}
	\caption{Visualization of a signotope on $23$ elements with no $6$-hole or $7$-gon as wiring diagram. 
 The triple orientations can be read as following: 
  $\orientp{a}{b}{c}$ with $a<b<c$ equals $+$ if and only if wire $a$ intersects $b$ before $c$ when traced from left to right.
 For more background on signotopes and wiring diagrams 
 see \cite{FelsnerWeil2001} and the handbook article~\cite{FelsnerGoodman2016}.}
	\label{fig:n23g7h6_wiring}
\end{figure}

We used SAT to show that every set of 30 points yields a 6-hole. Since there exist sets of 29 points \cite{Overmars2002} with no 6-holes, we determined the precise value $h(6)=30$. 
For Theorem~\ref{thm:h6g7=24} 
we do not have such a witnessing point set.
The SAT solver found millions of
signotopes on 23 elements
with no 7-gon and no 6-hole, witnessing that the bound 
is sharp in the more general combinatorial setting. 
Fig.~\ref{fig:n23g7h6_wiring} shows one such example.
However, so far we did not manage to find a corresponding point set to any of the signotopes. In fact, all tested configurations are provably non-realizable using the method of bi-quadratic final polynomials \cite{BokowskiRichter1990},
which is not surprising since 
only a small proportion ($2^{\Theta(n \log n)}$ of $2^{\Theta(n^2)}$) of rank $3$ signotopes are actually realizable by point sets; see  
\cite[Chapters~7.4 and~8.7]{BjoenerLVWSZ1993}. 
Moreover, 
deciding whether a triple-assignment can be realized by an actual point set is a notoriously hard problem as it is complete for the \emph{existential theory of the reals} ($\ETR$); 
a complexity class which lies between $\NP$ and $\PSPACE$~\cite[Chapter~8.4]{BjoenerLVWSZ1993}.

\fi

\end{document}